%% file: syntheticHammerheads.tex
\documentclass[letterhead,center,fleqn]{article}
\usepackage{xspace}
\usepackage{url}
\usepackage{graphicx}
\usepackage{amsmath}
\usepackage{amssymb}

\usepackage[margin=1in,nohead,centering]{geometry}
\newcommand{\rnaifold}{\mbox{\tt RNAiFold}\xspace}
\newcommand{\seq}{{\bf s}\xspace}

\title{Complete RNA inverse folding: computational design of functional
hammerhead ribozymes}

\author{Ivan Dotu\,$^{1,}$\footnote{The first three authors contributed equally to the work reported in this paper.},
Juan Antonio Garcia-Martin\,$^{1,\ast}$,
Betty L. Slinger\,$^{1,\ast}$,
Vinodh Mechery\,$^{2}$, \\
Michelle M. Meyer\,$^{1}$,
Peter Clote\,$^{1}$
\footnote{To whom correspondence should be addressed.
Tel: +1 617 552 1332; Fax: +1 617 552 2011; Email: clote@bc.edu}}
\date{%
$^{1}$Biology Department, Boston College, 140 Commonwealth Avenue,
Chestnut Hill, MA 02467 (USA) and
$^{2}$Hofstra North Shore-LIJ School of Medicine, Hempstead, NY
11549 (USA).
}

\begin{document}
\maketitle

\begin{abstract}
Nanotechnology and synthetic biology currently constitute one of the most
innovative, interdisciplinary fields of research, poised to radically
transform society in the 21st century.
This paper concerns the synthetic design of ribonucleic acid molecules, using
our recent algorithm, {\tt RNAiFold}, which can determine all RNA
sequences whose minimum free energy secondary structure is a 
user-specified target structure. 
Using {\tt RNAiFold}, we design ten {\em cis}-cleaving
hammerhead ribozymes, all of which are shown to be functional by
a cleavage assay. 
We additionally use {\tt RNAiFold} to design a functional 
{\em cis}-cleaving hammerhead as a modular unit of a synthetic larger RNA.
Analysis of kinetics on this small set of hammerheads suggests
that cleavage rate of computationally designed ribozymes 
may be correlated with positional entropy, ensemble defect, 
structural flexibility/rigidity and related measures. 

Artificial ribozymes have been designed in the past either manually or
by SELEX (Systematic Evolution of Ligands by Exponential Enrichment);
however, this appears to be the first purely computational design 
and experimental validation of novel functional ribozymes.
%{\rnaifold} is available via a web server at
{\rnaifold} is available at
{\small
\url{http://bioinformatics.bc.edu/clotelab/RNAiFold/}.
}
\end{abstract}

\section{Introduction}

Ribonucleic acid enzymes (a.k.a. ribozymes) 
are catalytic RNAs with enzymatic capabilities that, similar
to their protein counterparts, 
can catalyze and accelerate the rate of biochemical reactions while 
maintaining a great specificity with respect to the substrate they 
act upon. In general, ribozymes can catalyze the transesterification of 
phosphodiester bonds, acting in {\em cis} by self-cleavage, or in
{\em trans} by cleaving other RNAs.
There exist different types of ribozymes, all 
with a very well defined tertiary structure:
group I introns -- self-splicing ribozymes, that were first 
observed for the intron of the nuclear $26$S rRNA gene 
in {\em Tetrahymena thermophila} \cite{Cech81, Kruger82}; 
group II introns -- self-splicing ribozymes, which produce
ligated exons and an excised intron-lariat 
as products of the splicing procedure \cite{Peebles86}; 
ribonuclease P (RNase P) -- a ubiquitous endoribonuclease
that processes the $5'$ end of precursor tRNA molecules,
producing 5$'$ phosphoester and 3$'$ OH termini \cite{Darr92}; 
and small self-cleaving pathogenic RNAs, such as hammerhead ribozymes 
\cite{hammerhead:Xraystructure,hammerhead:structuralBasisOfCleavage},
as well as the hairpin and the hepatitis delta virus ribozymes 
\cite{Wilson.bst05}.\newline

\noindent
{\bf RNA Synthetic Biology}\newline

\noindent
In response to the increased understanding and appreciation of the role RNA plays in biology, the last decade has seen a surge 
in the field of RNA synthetic biology. Several laboratories have successfully produced
synthetic RNA sequences capable of self-cleaving, sensing small molecules {\em in vivo} or 
{\em in vitro}, as well as regulating gene expression
\cite{Isaacs06,Collins12}.  Many of these efforts have focused on the creation of allosteric ribozymes, or gene regulatory elements that can be used for further application.

Selection-based approaches (e.g. SELEX, or Systematic Evolution of 
Ligands by EXponential enrichment \cite{Ellington90,Tuerk90}) have proved
very powerful for generating a range of RNAs with a variety of capabilities.  Allosteric ribozymes that 
are inhibited or activated by specific small molecules 
have been achieved by utilizing a pre-existing self-cleaving ribozyme sequence coupled to either an existing aptamer \cite{Gu12}, or one derived through selection \cite{Piganeau09}. Additionally, SELEX has been coupled with {\em in vivo} screens to create RNAs with gene-regulatory activity in response to specific small molecule \cite{Sinha2010} or protein stimuli \cite{Goldfless2012, Belmont2010}.

% **************************************************************
% Keep this command to avoid text of first page running into the
% first page footnotes
\enlargethispage{-65.1pt}
% **************************************************************

Design-based approaches have also been successful at creating RNAs with 
engineered functions.  By a series of manually determined pointwise mutations,
where biological activity was repeatedly assayed for intermediate structures,
a single RNA sequence was designed to simultaneously
support the catalytic activities of both the self-cleaving
hepatitis delta virus ribozyme, and the class III self-ligating ribozyme
\cite{Schultes.s00}.
%SchultesBartel sequence:
%GAACCAGUCGGACUCCCAUUAGACUGGGCCGCCUCCUCGCGGCGGGAGUUGGGCUAGGGAGGAACAGCCUUUUCUAGGCUAAGGCCCA
Several approaches to designing genetic regulators mimic the 
action of small regulatory RNAs by 
introducing engineered trans-acting RNAs to
occlude a ribosome binding site or start codon to inhibit translation. 
Gene expression may be altered in such systems by inhibiting the original RNA 
with a second trans-acting RNA \cite{Isaacs04}, or through utilization of a ligand 
binding domain (aptamer) to induce an alternative RNA structure that does not interact 
with the transcript of interest \cite{Bayer05}.
In addition, hammerhead ribozymes have been used to target the HIV 
virus \cite{Zhou94,Bauer97} by modifying sequences within base-pairing regions to 
target a specific sequence of viral RNA. 

As the complexity of synthetic RNA devices increases, there is an 
increasing need to go beyond {\em ad hoc} manual approaches, and {\em in vitro}
selection methods.
RNA molecules have been rationally designed by the assembly of structural RNA
tertiary fragments/motifs, extracted from X-ray and NMR structures of natural
RNA molecules \cite{Shapiro.mmb08,Bindewald.jmgm08}; see also \cite{Afonin.np11}.
Using computational methods with {\em reaction graphs}, with subsequent validation
using atomic force microscopy,
molecular programs have been executed for a variety of dynamic DNA constructs,
ranging from hairpins, binary molecular {\em trees}, to bipedal walkers
\cite{Yin.n08}.  
RNA thermoswitches have been computationally designed and synthesized, that
are as efficient as natural thermoswitches, by applying 
the program, {\tt switch.pl} \cite{Flamm:01a}, which attempts to
minimize the following cost function for input RNA sequence
${\bf a} = a_1,\ldots,a_n$:
\begin{flalign*}
&(E_{T_1}({\bf a},S_1)-G_{T_1}({\bf a}) )
+( E_{T_2}({\bf a},S_2) -G_{T_2}({\bf a}) )\\
&-\xi((E_{T_1}({\bf a},S_1)-E_{T_1}({\bf a},S_2))\\
&+ (E_{T_2}({\bf a},S_2)-E_{T_2}({\bf a},S_2))
\end{flalign*}
where $G_T({\bf a})$ is the ensemble free energy sequence  ${\bf a}$
at temperature $T$,  $E_T({\bf a},S)$ is the free energy of  RNA
sequence ${\bf a}$ with structure $S$
at temperature $T$,  and $0< \xi < 1$ is a constant. 
Waldminghaus et al. \cite{Waldminghaus08} selected promising thermoswitch
candidate sequences returned by {\tt switch.pl} by considering the
cost function values, the predicted melting temperature 
({\tt RNAheat} \cite{Gruber08}) etc. The resulting candidates were
not functional; however, functional thermoswitches were obtained
from these candidates after several rounds of 
error-prone PCR mutagenesis and in vitro selection.
Recently, a synthetic theophylline riboswitch has been
rationally designed to {\em transcriptionally} regulate the expression of
a gene, by fusing a theophylline aptamer with a computationally designed
expression platform \cite{Wachsmuth.nar13}.
However, to the best of our knowledge, no group has previously
designed a ribozyme by
purely computational means, using RNA inverse folding, and subsequently 
validated the ribozyme functionality; this is our contribution in the
present article. \newline

\noindent
{\bf Molecular Design and RNA Inverse Folding} \newline

\noindent
Given an RNA sequence, the {\em folding} problem is to determine
the {\em native structure} into which the sequence folds; in contrast,
given a target RNA structure, the {\em inverse folding} problem is to
determine one, several, or all sequences whose native structure is the
given target structure.  Since the pioneering
work of Anfinsen \cite{anfinsen}, it is widely accepted that the native 
structure of a given macromolecule can be identified with its
minimum free energy (MFE) structure. 
If we identify native structure with the MFE tertiary structure, then both the
folding and inverse folding problems are NP-complete
\cite{bergerLeighton,Lyngso.jcb00}. However, since
RNA secondary structure appears to form prior to 
tertiary interactions, thus creating a scaffold for tertiary
structure formation \cite{banerjeeJaegerTurner,Bailor.s10}, and since
the folding and inverse folding problems are intractable for tertiary
structures, we consider the folding and inverse folding 
problems for RNA secondary structure in this paper.

Using free energy parameters obtained from optical melting
experiments \cite{turner}, the dynamic programming algorithm of Zuker
\cite{zukerStiegler} determines the MFE secondary structure of a given
RNA in cubic time. This algorithm has been implemented in
%{\tt mfold} \cite{zuker:mfoldWebserver}, 
%{\tt UNAFOLD} \cite{Markham.mmb08}, 
%Vienna RNA Package {\tt RNAfold} \cite{hofacker:ViennaWebServer,Gruber08},
%and {\tt RNAstructure} \cite{mathews:RNAstructure}, 
\cite{zuker:mfoldWebserver,Markham.mmb08,hofacker:ViennaWebServer,Gruber08,mathews:RNAstructure,Lorenz.amb11},
where it should be noted
that secondary structure predictions may differ due to different
treatment of dangles, coaxial stacking, etc. and their corresponding energy 
parameters.

It seems likely that inverse folding is NP-complete, even for RNA
secondary structures \cite{schnallLevinInverseViterbi}; nevertheless,
a number of heuristic algorithms exist that return approximate
solutions:
{\tt RNAinverse} \cite{hofacker:FastFolding},
{\tt switch.pl} \cite{Flamm:01a},
{\tt RNA-SSD} \cite{Andronescu04},
{\tt INFO-RNA} \cite{Busch06},
{\tt MODENA} \cite{Taneda11}, 
{\tt NUPACK-DESIGN} \cite{Zadeh.jcc11},
{\tt Inv} \cite{Gao10},
{\tt Frnakenstein} \cite{Frna}.
Rather than employing a heuristic, our
recent software, {\tt RNAiFold} \cite{GarciaMartin13,GarciaMartin.nar13},
employs Constraint Programming (CP) \cite{pascal:CometBook}, which
always returns exact solutions,  although it might do so in an impractical 
amount of time. Moreover, CP is the only inverse folding software
capable of determining whether (provably) no solution exists --
i.e., that no RNA sequence has MFE secondary structure that is identical to 
the target structure.
Additionally, CP allows us to model and account for several 
RNA sequence design constraints that are necessary for a more biologically
relevant result -- for instance, controlling GC content, 
describing fixed upper and lower bounds for certain types of base pairs,
limiting a maximum number of consecutive nucleotides of a given type,
specifying certain mononucleotide and/or dinucleotide frequencies,
requiring specific nucleotides that are suspected to constitute the active
site, etc. CP can also enforce {\em compatibility} constraints and 
{\em incompatibility} constraints, which require that 
all returned sequences not only fold into the given target structure,
but additionally are compatible (incompatible)
with another user-stipulated structure.

\section{MATERIALS AND METHODS}

\subsection{Computational Methods}

{\tt RNAiFold} returns sequences whose minimum free energy (MFE)
structure is a given target structure, whereby the user may choose to use
the free energy parameters from either Vienna RNA Package 1.8.5 (Turner 1999
parameters) or Vienna RNA Package 2.0.7
(Turner 2004 parameters) \cite{Turner.nar10}.  By abuse of notation, let
{\tt RNAiFold} 1.8.5 [resp. 2.0.7] denote the program {\tt RNAiFold} with
energy parameters from the corresponding version of Vienna RNA Package.

As target structure for our computationally designed
type III hammerheads, we selected the secondary structure of a portion
of the plus polarity strand of Peach Latent Mosaic Viroid (PLMVd)
(isolate LS35, variant ls16b) from Rfam family RF00008
\cite{Gardner.nar11} having accession code AJ005312.1/282-335.
The reason we chose PLMVd AJ005312.1/282-335
was that this is the only RNA sequence in the
seed alignment of RF00008, whose MFE structure is identical to its Rfam
consensus structure, when computed by
{\tt RNAfold} 1.8.5 -- see Supplementary Information (SI) for a precise
definition of Rfam consensus structure. Moreover, as shown in SI Figure 1,
the MFE structure computed by {\tt RNAfold} 2.0.7 differs markedly from
the Rfam consensus structure of PLMVd AJ005312.1/282-335, hence we
used {\tt RNAiFold}
with the energy parameters from Vienna RNA Package 1.8.5.
In summary, the target structure for {\tt RNAiFold} 1.8.5 was taken to be
\begin{quote}
\begin{tiny}
%GAUGAGUCUGUGCUAAGCACACUGAUGAGUCUAUGAAAUGAGACGAAACUCAUA
\begin{verbatim}
.((((((.(((((...))))).......((((........))))...)))))).
\end{verbatim}
\end{tiny}
\end{quote}
which is both the {\tt RNAfold} 1.8.5 MFE structure as well as the
Rfam consensus structure of PLMVd AJ005312.1/282-335.

Numerous biochemical and structural studies have pinpointed key
nucleotides in the hammerhead ribozome that are required for
catalysis \cite{Blount.arbb05,Martick.c06,Nelson.r08}. However for
an efficient, purely computational design of synthetic hammerheads,
it is important to rely only on sequence conservation results from
reliable multiple alignments. The Rfam web site image
\url{http://rfam.sanger.ac.uk/family/RF00008\#tabview=tab3}
clearly shows certain regions of the 56 nt consensus sequence have
highly conserved sequence identity. Based on this observation, we
computed the nucleotide frequency for the seed alignment of Rfam
family RF00008 for those positions aligned to the nucleotides of
the 54 nt PLMVd  AJ005312.1/282-335.
%Figure~\ref{fig:PLMVdSequenceLogo} (left) shows the
Figure~1 (left) shows the
sequence logo of positions aligned to 
PLMVd AJ005312.1/282-335. 
%Table~\ref{table:rankConservedPositionsPLMVd} shows that 
SI Table~1 shows that 
sequence identity exceeds 96\% for the 
15 positions 6-7, 22-25,27-29, 44-49 of PLMVd in the
seed alignment for Rfam family RF00008 consisting of
84 sequences.
For that reason, the nucleotides in PLMVd at these 15 
positions were provided as a constraint for {\tt RNAiFold},
thus fixing approximately 28\% of the 54 nucleotides.
Note that the cleavage site at C8, discussed
below would have been included in the constraints, had we chosen to retain
positions of at least 95\%.

From the literature, it is well-known that hammerhead cleavage sites
are of the form NUH (e.g. GUH and CUH); see, for instance, papers of
Pan et al.  \cite{hammerheadCleavageSite} and
Gonzalez-Carmona et al.  \cite{hammerheadCleavageSiteBis}, which provide
experimental data on the efficiency of various target hammerhead
cleavage sites. For PLMVd, cleavage occurs immediately after the cytidine
at position 8.  For this reason, IUPAC code H (i.e. not G) was given as an
additional constraint at position 8 for {\tt RNAiFold}. 

Apart from nucleotide constraints at positions 6-7, 22-25,27-29, 44-49, 
and the constraint H8, 
all nucleotides at the remaining 38 positions were constrained 
to be {\em distinct} from those of PLMVd -- this was done to prevent any
unintentional use of other nucleotide identities in the computational
design of a hammerhead.  Summarizing, 
each sequence returned by {\tt RNAiFold} was required to satisfy
IUPAC sequence constraints given by
{\tt HBVHBGUHVH VHDVBBHDBD BCUGAVGAGV DVBVHBBBVH BHBCGAAACV DBVB}
as shown in
%Figure~\ref{fig:PLMVdSequenceLogo} (right);
Figure~1 (right);
moreover, the MFE structure of each returned sequence, determined by
{\tt RNAfold} 1.8.5, is necessarily identical to the target
consensus structure of PLMVd, as shown in
%Figure~\ref{fig:superimposedBinaryFullEntropyPlotHHwildtype}
Figure~2.

{\tt RNAiFold} was run four times, each time additionally
constraining GC content
to be within a specified range. Altogether, over one million solutions of
RNA inverse folding were returned before memory exhaustion (using the 32 bit
version of run-time system {\tt COMET}):
200,072 with GC-content 30-39\%,
352,924 with GC-content 40-49\%,
349,325 with GC-content 50-59\%,
366,323 with GC-content 60-69\%,
constituting a total of 1,268,644 sequences. Output 
sequences $\seq$ were selected according to a number of criteria
explained below.

Measures used in selecting promising hammerhead candidates from
{\tt RNAiFold} were of two basic types that addressed the following
questions: {\em (1)} To what extent
do low energy structures of $\seq$ resemble the
%target Rfam consensus structure of PLMVd AJ005312.1/282-335?
MFE structure?
{\em (2)} To what extent are the same
structural regions of PLMVd AJ005312.1/282-335 
as {\em flexible/rigid} as those 
of $\seq$? In other words, the measures used for sequence selection concern
either {\em structural diversity} or regional {\em structural 
flexibility/rigidity}; in particular, 
no sequence homology measures were used in
selecting candidate hammerhead sequences for testing, including
the program {\tt Infernal} \cite{Nawrocki.b09}.

One measure of type 1 is the Boltzmann probability $P(S_0,\seq)$, 
where $S_0$ denotes the MFE structure of $\seq$ (identical to the 
Rfam consensus structure of PLMVd AJ005312.1/282-335, since {\tt RNAiFold}
solves inverse folding), and 
$P(S_0,\seq) = \frac{\exp(-E(S_0,\seq)/RT}{Z}$, where $E(S_0,\seq)$ is the
free energy of structure $S_0$ for sequence $\seq$, as computed by
Turner 1999 energies, and $Z$ is the partition function.
Other measures of type 1 are
average structural positional entropy \cite{Huynen.jmb97},
ensemble defect \cite{Dirks.nar04},
expected base pair distance \cite{GarciaMartin13},
Vienna structural diversity \cite{Gruber08}, 
Morgan-Higgs structural diversity \cite{higgs:sampling}.
Additionally, the restriction of these measures to
the positions 6-8, 22-25, 27-29, 44-49, was computed. Throughout this
paper, 
we use the term {\em conserved site} to denote these 16
positions (we use the term {\em conserved site}, rather than
{\em active site}, which has a different meaning in
the biochemical literature). Thus we included measures such as
average (structural positional) entropy of conserved site,
ensemble defect of conserved site, etc. Measures of type 2 concern
the maximum discrepancy between values of type 1 for a candidate
sequence $\seq$ and wild type PLMVd AJ005312.1/282-335. These are
briefly explained in the next section;
see \cite{GarciaMartin13,GarciaMartin.nar13} or Supplementary
Information.
\newline

\noindent
{\bf Structural positional entropy:}
In selecting the most promising candidate hammerheads from the sequences
returned by {\tt RNAiFold}, we additionally considered 
{\em discrepancy} (deviation) from
structural positional entropy of conserved positions in PLMVd. Unlike
the notion of nucleotide positional entropy used in sequence logos
\cite{Crooks.gr04}, structural positional entropy is defined as follows.
If $n$ is the length of a given RNA sequence,
then for $1 \leq i,j \leq n$, let $p^*_{i,j}$ denote
the probability $p_{i,j}$ of base pair $(i,j)$ if $i<j$,
the probability $p_{j,i}$ of base pair $(j,i)$ if $j<i$,
and the probability that $i$ is unpaired, $i=j$.  With
this notation, the (structural) entropy of position $i$ is defined by 
$H(i) = - \sum_{j} \left( p^*_{i,j} \log p^*_{i,j}
+ (1-p^*_{i,j}) \log (1-p^*_{i,j}) \right)$.  Base 2 logarithms are
usually used, whereby entropy is given in bits, ranging from
a minimum value of $0$, where $p^*_{i,j_0}=1$ for some $j_0$,
to a maximum value of $\ln n/\ln 2$, 
in the case that $p^*_{i,j}=1/n$ for each $j$.  

An alternative to (full) structural positional entropy is binary 
structural positional
entropy, defined by 
$H_b(i) = -\left( p^*_{i,i} \log p^*_{i,i} + (1-p^*_{i,i}) 
\log( 1-p^*_{i,i})\right)$. 
Binary positional entropy values $H_b(i)$ range from 
a minimum value of $0$ bits,
where position $i$ is either always base paired (though possibly to distinct
partners) or always unpaired in the low energy ensemble of structures, to
a maximum value of $1$, where position $i$ is paired (unpaired) with exactly
probability $1/2$. 
%Figure~\ref{fig:superimposedBinaryFullEntropyPlotHHwildtype}
Figure~2 displays full and binary structural
positional entropy for PLMVd AJ005312.1/282-335. 

At the 16 conserved positions 6-8, 22-25, 27-29, 44-49 of PLMVd, 
there is a range of structural positional entropy values, suggesting
that certain nucleotides may be located within a more flexible
(high entropy) region of the structure, while other nucleotides
may be located within a more rigid (low entropy) region. 
%Figure~\ref{fig:superimposedBinaryFullEntropyPlotHHwildtype} indicates the
Figure~2 indicates the
structural entropy of nucleotides within the consensus structure of PLMVd
by appropriate colors, as well as a function of position.

Hypothesizing that low [resp. high] entropy regions of the 
hammerhead ribozyme could indicate structural {\em rigidity}
[resp. {\em flexibility}] requirements
necessary for hammerhead function, we scrutinized the sequences
returned by {\tt RNAiFold} by measures of {\em deviation (or discrepancy) 
from structural
positional entropy} of PLMVd AJ005312.1/282-335. This led to a number of
measures, formally defined in Section 1 of the 
Supplementary Information (SI), of which
the most important are the following:
full/binary entropy discrepancy for complete sequence
defined in SI equations (7) and (8),
full/binary entropy discrepancy for the conserved site
defined in SI equations (20) and (21)
(recall that `conserved site' denotes
the 16 positions 6-8, 22-25, 27-29, 44-49  constrained
by {\tt RNAiFold}).
{\em Entropy discrepancy} for the complete sequence [resp. conserved site]
is defined to be the
maximum, taken over all 54 positions [resp. over positions
6-8, 22-25, 27-29, 44-49], of the absolute value of the difference between
structural entropy of a candidate returned by {\tt RNAiFold} and
that of PLMVd. 
\newline

\noindent
{\bf Sequences selected:}
%Table~\ref{table:tableSequencesChosen} 
Table~1
shows the candidate hammerhead
sequences finally selected for cleavage assay, together with
the selection criteria used for each sequence.
Ten candidate hammerheads were selected: HH1-HH10.
HH1-HH5 were chosen from sequences of specific GC-content ranges, 
to have have the smallest {\em binary} entropy discrepancy for
the `conserved site'.
%the `conserved site', consisting of
%16 conserved positions 6-8, 22-25,27-29, 44-49.
HH1 was selected from sequences having GC-content 30-39\%;
HH2 from sequences having GC-content 40-49\%;
HH4 from sequences having GC-content 50-59\%;
HH5 from sequences having GC-content 60-69\%.
Since PLMVd AJ005312.1/282-335 has GC-content of 40.7\%, 
HH3 was chosen to have second smallest {\em binary} entropy distance 
for the conserved site, selected from sequences having GC-content 40-49\%.

Additional candidate hammerheads were chosen by different criteria,
in order to determine their effect on functionality.
HH6 was chosen to have the {\em largest} {\em binary} entropy discrepancy
for the conserved site, selected from all sequences having C at cleavage
position 8, provided that the Boltzmann probability of the MFE structure
exceeded $40\%$.
HH7 was chosen to have the {\em smallest}
ensemble defect of all sequences having C at cleavage position 8.
HH8 was chosen to have the smallest ensemble defect of all sequences, 
regardless of nucleotide at position 8 (HH8 has A at cleavage site,
instead of C).
HH9 was chosen to have the smallest {\em binary} entropy discrepancy
for the `conserved site', selected from all sequences, for which the
probability $P(S_0,\seq)$ of the target PLMVd structure was {\em at most}
$0.2$.  Finally,
HH10 was chosen to have the smallest {\em binary} entropy discrepancy
for the conserved site, selected from all sequences, regardless of
probability of target structure.
Note that HH1-HH6 were selected with the requirement that
$P(S_0,\seq) \geq 0.4$, while HH7-HH10 were selected without this
requirement.
This was done in order to determine how important target structure probability
might be in hammerhead functionality.
\newline

\noindent
{\bf Computational Pipeline Summary:}
The following computational pipeline summarizes the generation and
selection of candidate hammerhead sequences. 
%\begin{quote}
\begin{tiny}
\begin{verbatim}
1. find Rfam sequence, whose MFE structure resembles family consensus structure
2. determine highly conserved positions in reliable multiple alignment
3. run RNAiFold to solve the constrained inverse folding problem
4. filter using Boltzmann probability, GC-content, entropy, ensemble defect, etc.
5. perform biochemical validation
\end{verbatim}
\end{tiny}
%\end{quote}
A Python program can be downloaded from the {\tt RNAiFold} web site, 
that automates steps 1,2. Of course, one can bypass step 1 without
using Rfam, and instead use any reliable multiple sequence/structure alignment.
\newline

\noindent
{\bf Design of modular hammerhead within another structure:}
It has many times been observed that aptamers, hammerheads and other 
functional RNAs constitute {\em modules}, capable of function even
when engineered to form part of a larger RNA molecule. 
For instance, Wieland et al.  \cite{Wieland.acie08} 
created artificial {\em aptazymes} 
by replacing a hammerhead helix by a theophylline aptamer, and
Saragliadis et al.  \cite{Saragliadis.rb13} created artificial
{\em thermozymes}, created by fusing
a theophylline aptamer to a {\em Salmonella} RNA thermometer
\cite{Saragliadis.rb13}.

With the intent of designing a guanine-activated
riboswitch with a modular hammerhead,
we followed the following steps in rationally designing a synthetic 166 nt
RNA, with putative type III hammerhead module.
Target secondary structure $S$ was taken to be the structure of the 
gene {\sc off} xanthine phosphoribosyltransferase (XPT) riboswitch, 
depicted in Figure 1A of \cite{Serganov.cb04},
whereby the terminator loop (expression platform) was replaced by
the Rfam consensus structure for a type III hammerhead. Sequence
constraints were chosen to be the highly conserved nucleotides of the 
Rfam consensus structures for the purine riboswitch (RF00167 seqcons
view of consensus structure) and for type III
hammerhead (RF00008 seqcons view of consensus structure). 
% \cite{Mandal.nsmb04}.
%Figure~\ref{fig:riboswitchRibozyme} displays the target structure $S$ for
Figure~3 displays the target structure $S$ for
computational design of a modular hammerhead within the terminal
stem-loop of a structure similar to the XPT riboswitch. 
We gave {\tt RNAiFold}
an additional compatibility constraint, whereby returned sequences 
were required to
be compatible to a second structure $S'$, in which the hammerhead
cleavage site (NUH) is fully sequestered within a base-paired region.
Positions 60-118 of $S'$ are given as follows:
%\begin{quote}
\begin{tiny}
\begin{verbatim}
5'-ACUAYNNNNNNNNNNNNNNNNNNNNNNNNNNNNNNNNNNNNNNNNNNNNNNNNNGUHNN-3' 
5'-.....(((((((((((..(((((........)))))............)))))))))))-3'
\end{verbatim}
\end{tiny}
%\end{quote}
while all positions in $S'$ outside of 60-118 (i.e. from 1-59 and
119-166) are unpaired.

We filtered sequences output by {\tt RNAiFold}, by applying
{\tt RNAbor} \cite{FreyhultMoultonClote:RNAbor}, and its faster sequel,
{\tt FFTbor} \cite{FFTbor}. Given reference structure $S$,
{\tt RNAbor} and {\tt FFTbor} return the {\em density of states} with 
respect to $S$, which depicts the Boltzmann probability
$p(k)=\frac{Z_k}{Z}$ for secondary structures to have base
pair distance $k$ from $S$.  Additionally, {\tt RNAbor}
computes, for each $k$, the MFE$_k$-structure; i.e. that structure 
having minimum free energy over all
structures whose base pair distance from the reference structure  $S$
is exactly $k$.  

From a partial output of 3,000 sequences from {\tt RNAiFold}, 
only one sequence $\seq$ satisfied the following two properties, 
when applying {\tt RNAbor} with input $\seq$ and reference MFE structure $S$:
$(1)$ The density of states figure has a pronounced peak at $k=0$, 
corresponding to the location of the MFE structure $S$;
$(2)$ There was another pronounced peak for value $k \gg 0$, corresponding
to a structure $T$ containing the base pairs in $S'$, which thus should
sequester the ribozyme cleavage site NUH, located at position
114-116 -- see SI Figure~3..  
 
The final, selected sequence 166 nt $\seq$ is given 
as follows:
{\tiny
GCCGC GUAUA AGGGC UGCGA UAAGG GCAGU CCGUU UCUAC GGGCG
GCCGU AAACC GCCCA CUACG CGGCG UGGUU AAGCC GGAAA GGAGA
CCGGC AGGAG GGUAA UGGGC CGCGU CGCGG CGCGG GAGCG CGCCG
CCUGA UGAGU CCGUG AGGAC GAAAC GCGGCC}.

\subsection{Experimental Validation}

Complementary DNA oligonucleotides, corresponding to the DNA sequence 
of the designed RNAs preceded by a T7 RNA polymerase promoter, were 
purchased from MWG Operon. The ten hammerhead candidate sequences
HH1-HH10, extended 2 nt on the left by GG and 2 nt on the right by CC
for transcriptional efficiency,
and the 166 nt sequence, harboring a candidate
hammerhead in the rightmost stem-loop of 
SI Figure~3 were constructed 
using primer extension and PCR amplified (5 U taq polymerase (New 
England Biolabs), 2.5 mM each NTP, 1x NEB Thermopol buffer). 
For each of the 10 designed hammerhead sequences, the H8G mutant 
was constructed in a similar manner, using alternative oligonucleotides 
containing the mutation. Similarly, C116G 
(analogous to H8G) and G142U mutations 
were constructed for the 166 nt designed ribozyme.
The resulting PCR products were TOPO-cloned 
(Invitrogen), and the designed and mutant sequences were 
verified by sequencing plasmids containing full-length PCR products. 
These plasmids were subsequently used as templates for PCR reactions to 
generate template for {\em in vitro} transcription. 

To generate the RNA, {\em in vitro} transcription was performed using T7 RNA 
polymerase (400 U T7 polymerase, 80 mM HEPES-KOH pH 7.5, 24 mM MgCl$_2$, 
2 mM spermidine, 40 mM DTT, 2 mM each NTP) with the addition of 10 
$\mu$Ci of $\alpha$-32P-GTP for transcriptions to generate body-labeled RNA 
when necessary. To prevent premature cleavage during
transcription, 100 uM of oligonucleotides complementary to nucleotides 17-35 (numbering starts after the leading GG) 
were added to each reaction. Full-length RNAs were purified using denaturing 
PAGE (20\% acrylamide).

To assess self-cleavage of designed hammerhead sequences, 
RNA was incubated for 1 hour in cleavage buffer (5 mM MgCl$_2$, 
50 mM tris pH 7.5) at 25$^{\circ}$C. 
Subsequently, 1 volume of 2x gel-loading buffer  (16 M urea (supersaturated), 
10 mM EDTA, 20\% sucrose, 0.1\%SDS, 100 mM tris pH 8.0, 100 mM borate, 
0.05\% bromophenol blue) was added to quench the reaction with final 
urea and EDTA concentrations of 8 M and 5 mM respectively.
The reaction was placed on ice until gel loading.

%Peter hours changed
%Samples lacking Mg$^{++}$ were incubated in 50 mM tris pH 7.5 for 1 hour at 25$ ^{\circ}$C.  For the 166 nt RNA, cleavage experiments were conducted under similar conditions but reactions were incubated for a few seconds (0 hours), 30 minutes, 5 hours and 24 hours and samples lacking Mg$^{++}$  were incubated in 50 mM tris pH 7.5 for 24 hours at 25$ ^{\circ}$C.  Cleavage products were separated by denaturing PAGE (10\% acrylamide), and the gels dried prior to exposure to phosphoimager plates (GE Healthcare) for 18 hours. The gels were imaged using a STORM 820 phospoimager (GE Healthcare).
Samples lacking Mg$^{++}$ were incubated in 50 mM tris pH 7.5 for 1 hour at 25$ ^{\circ}$C.  For the 166 nt RNA, cleavage experiments were conducted under similar conditions but reactions were incubated for a few seconds (0 h), 30 min, 5 h and 24 h, and samples lacking Mg$^{++}$  were incubated in 50 mM tris pH 7.5 for 24 h at 25$ ^{\circ}$C.  Cleavage products were separated by denaturing PAGE (10\% acrylamide), and the gels dried prior to exposure to phosphoimager plates (GE Healthcare) for 18 h. The gels were imaged using a STORM 820 phospoimager (GE Healthcare).
\newline

\noindent
{\bf Kinetics:} 
To determine the cleavage rates for designed hammerhead sequences, body-labeled RNA
was incubated in cleavage assays as described above for varying amounts of 
time. 
%A typical 
%time series for a hammerhead with a cleavage rate of approximately 0.01/min 
%includes time points at 0 min,  30 seconds, 1 minute, 10 minutes, 
%30 minutes, 2 hours, 4 hours, 6 hours, 24 hours and 48 hours. 
%For faster hammerheads with a cleavage rates in the range of 0.1/min the 
%time series was shifted  to include time points at 0 min, 
%30 seconds, 1 minute, 
%5 minutes, 10 minutes, 30 minutes, 2 hours, 6 hours and 24 hours. 
Cleavage products were separated 
and gels imaged as described above. The cleavage products were quantified using 
ImageQuant software (GE Healthcare). To calculate the fraction cleaved at time $t$, $F(t)$,
the sum of the quantified counts for 5$'$ and 3$'$ cleavage 
product bands was divided by the total quantified counts for the 
entire reaction (uncleaved, 5$'$ and 3$'$ cleavage products).  

The observed cleavage rate 
$K_{\mbox\small obs}$ was computed by using the {\tt Matlab} function
{\tt nlinfit} with constant error model to fit cleavage time series
data using the equation
\begin{eqnarray}
\label{eqn:1}
%\frac{F_{\mbox\small \max}-F(t)} {F_{\mbox\small \max}-F(0)} = \exp(
%K_{\mbox\small obs}\cdot t)
F_{\mbox\small \max}-F(t) = (F_{\mbox\small \max}-F(0)) \cdot \exp(
K_{\mbox\small obs}\cdot t)
\end{eqnarray}
where $F(t)$ denotes the amount of cleavage product measured at time $t$,
and $F_{\mbox\small \max}$ the maximal fraction cleaved. 
The 95\% confidence interval of this fit was calculated from the 
resulting residuals and variance-covariance matrix using the 
Matlab function {\tt nlpredci}.  
See SI Figure 2.
\newline

%\noindent
%{\bf Kinetics:} The observed cleavage rate 
%$K_{\mbox\small obs}$ was computed by using the {\tt Matlab} function
%{\tt nlinfit} with constant error model to fit cleavage time series
%data using the equation
%\begin{eqnarray}
%\label{eqn:1}
%%\frac{F_{\mbox\small \max}-F(t)} {F_{\mbox\small \max}-F(0)} = \exp(
%%K_{\mbox\small obs}\cdot t)
%F_{\mbox\small \max}-F(t) = (F_{\mbox\small \max}-F(0)) \cdot \exp(
%K_{\mbox\small obs}\cdot t)
%\end{eqnarray}
%where $F(t)$ denotes the amount of cleavage product measured at time $t$,
%and $F_{\mbox\small \max}$ was determined after 48 hours.
%The 95\% confidence interval of this fit was calculated from the 
%resulting residuals and variance-covariance matrix using the 
%Matlab function {\tt nlpredci}.
%
%To determine the kinetics of a  hammerhead having cleavage rate of the
%order of magnitude of roughly 0.01/min, a typical time series was the following:
%0 min, 10 seconds, 1 minute, 10 minutes, 30 minutes, 2 hours,
%4 hours, 6 hours, 24 hours, 48 hours.
%For faster hammerheads (on the order of 0.1/min), the time series typically
%the following:
%0 min, 10 seconds, 30 seconds, 1 min, 5 minutes,
%10 minutes, 30 minutes, 2 hours, 6 hours, 24 hours, 48 hours.
%See Supplemental Information for cleavage rate curves.

\subsection{RESULTS} 
%\subsection*{Computational Results}

Given the target Rfam consensus structure $S$
of Peach Latent Mosaic Viroid (PLMVd) AJ005312.1/282-335, which
is identical with the MFE secondary structure using {\tt RNAfold} 1.8.5,
16 highly conserved positions nucleotides were taken as constraints
in the generation of over one million sequences solving the inverse
folding problem, as determined by {\tt RNAiFold} 1.8.5.
Using distance measures of {\em dissimilarity} of low energy
structures to the MFE structure (positional entropy, ensemble defect, 
structural diversity, etc.)
together with measures of molecular structural flexibility/rigidity,
ten putative hammerhead sequences were selected for {\em in vitro}
validation using a cleavage assay. The selected sequences and
selection criteria are given in
%Table~\ref{table:tableSequencesChosen}
Table~1.  All ten hammerhead candidates, listed in this table,
were shown to be functional, with cleavage rates
%, as computed by equation~(\ref{eqn:1}), appearing in 
listed in 
%Table~\ref{table:kinetics}. These values should be compared with
Table~2. 
Cleavage assay gel images for the designed hammerheads HH1-HH10 are displayed
%in Figure~\ref{fig:cleavageHH1-HH10}, 
in Figure~4,  where each sequence shows Mg$^{++}$-dependent cleavage.
In addition, the H8G mutant of each designed hammerhead shows no activity.
These data strongly suggest that  the designed sequences HH1-HH10 
behave in a manner consistent with the expected mechanism for 
hammerhead ribozymes. 
Time series for cleavage fraction and kinetics curves for a typical 
designed hammerhead ribozyme (HH1) and the fastest designed 
ribozyme (HH7) are shown in 
%Figure~\ref{fig:kineticsHH1_HH7},
Figure 5, 
while similar figures for the remaining designed hammerheads 
appear in SI Figure 2.
Kinetics for the designed hammerheads should be compared with
wild type hammerhead kinetics, where
under standard conditions of 10 mM MgCl$_2$, pH 7.5, and 25$^{\circ}$ C,
cleavage rates between 0.5 and 2 per minute have been observed
for at least 20 different hammerheads \cite{ClouetdOrval.b97}.
It follows that kinetics of the
computationally designed hammerheads described in this paper are slower
than wild type hammerheads approximately by a factor of 10.

Pearson correlation coefficient was determined between cleavage rate
$K_{\mbox\small obs}$, obtained by fitting equation (\ref{eqn:1})
with data from three to five 
technical relicates, and 21 measures, including average positional 
entropy, GC-content, minimum free energy, etc. See Supplementary 
Information for all correlation values. The most pronounced
correlations were observed between $K_{\mbox\small obs}$
and (full) average structural positional entropy,
ensemble defect, and expected base pair distance
discrepancy for `conserved site' with values respectively of
-0.461, -0.370, -0.438; i.e. cleavage is faster when these measures
are smaller. See SI equations (7),(5) and (22) for formal definitions 
of these notions.

%It is known from literature \cite{hammerheadCleavageSite,hammerheadCleavageSiteBis}
%that hammerhead cleavage sites are of the form NUH (e.g. GUH and CUH,
%but not GUG), although, to the best of our knowledge, there is no
%published explanation for this phenomenon.  %Peter added 'published'
%%in Figure~\ref{fig:cleavageHH1-HH10}, 
%Figure~4 shows that
%the H8G mutant of each designed sequence HH1-HH10 
%does not cleave under mild denaturing conditions that suffice
%for cleavage of HH1-HH10.
%In addition,
%{\tt RNAiFold} determined that (provably) there is no RNA sequence,
%whose MFE structure is the Rfam consensus structure of
%Peach Latent Mosaic Viroid (PLMVd) AJ005312.1/282-335, 
%having a guanine at cleavage site 8, as well as the 15 highly conserved
%nucleotides of PLMVd at positions 6-7, 22-25,27-29, 44-49
%(left panel of 
%%Figure~\ref{fig:nonCleavingStructures}.
%Figure~6). This result holds for both the Turner 99 and Turner 2004
%energy models. 

It is known from 
literature \cite{hammerheadCleavageSite,hammerheadCleavageSiteBis}
that hammerhead cleavage sites are of the form NUH (e.g. GUH and CUH,
but not GUG). Indeed, Carbonell et al. \cite{Carbonell.nar06} 
%right column, page 5620
suggest that G8 would pair with C22 (in our numbering) and impede
its role in the catalytic pocket.
Figure~4 shows that the H8G mutant of each designed sequence HH1-HH10 
does not cleave under mild denaturing conditions that suffice
for cleavage of HH1-HH10.  In addition,
{\tt RNAiFold} determined that (provably) there is no RNA sequence,
whose MFE structure is the Rfam consensus structure of
Peach Latent Mosaic Viroid (PLMVd) AJ005312.1/282-335, 
having a guanine at cleavage site 8, as well as the 15 highly conserved
nucleotides of PLMVd at positions 6-7, 22-25,27-29, 44-49 (left panel of 
%Figure~\ref{fig:nonCleavingStructures}.
Figure~6). This result holds for both the Turner 99 and Turner 2004
energy models.

Since {\tt RNAiFold} also solves the inverse hybridization problem, we 
considered the NUH cleavage target of {\em trans}-cleaving
hammerhead ribozymes, known from comparative sequence analysis \cite{James11}.
Application of {\tt RNAiFold} showed that there do not exist any two sequences,
where the first contains GUG at the
cleavage site location, for which the minimum free energy hybridization
structure is the target structure appearing in
the right panel of 
%Figure~\ref{fig:nonCleavingStructures}.
Figure~6. Taken together, these results provide a compelling
computational explanation for the reason that
GUG is not a hammerhead cleavage site.

To demonstrate the functionality of a computationally designed
hammerhead, occurring within a larger rationally designed RNA,
we synthesized the 166 nt sequence $\seq$, 
designated as `synthetic wild type', 
as well as two mutant sequences $\seq_1$, $\seq_2$, each containing
a mutation that should inactivate hammerhead activity. Sequence $\seq_1$
contains a C116G mutation at the GUC site of cleavage, while $\seq_2$
contains a G142U mutation in a distal section 
of the ribozyme, known to be required for cleavage (the CUGAUGA sequence). 
Cleavage 
assays under mild conditions (5 mM MgCl$_2$, 50 mM tris pH 7.5, 25$^{\circ}$C) 
show that approximately 40\% of our synthetic 
wild type sequence rapidly cleaves 
at the expected site (see  Supplementary  
Information Figure 4  for T1 mapping of the cleavage products),
in the absence and presence of guanine. 

The cleavage is Mg$^{++}$-dependent 
%(Figure~\ref{fig:ribozymeCleavageAssay}A),
(Figure 7A), and the hammerhead appears to cleave rapidly within seconds. 
Neither of the mutant sequences displays any cleavage under the same 
conditions, even with significantly longer incubation times 
%(Figure~\ref{fig:ribozymeCleavageAssay}B,C),
(Figure 7B,C).  Kinetics for the 166 nt synthetic ribozyme are 
comparable with those of wild type hammerheads, with an observed 
cleavage rate $K_{\mbox\small obs}$ of 1.3/min and Fmax of 0.47 
%(Figure~\ref{fig:ribozymeCleavageAssay}D).
(Figure 7D). Addition of 1 mM guanine has no significant affect on 
either the $K_{\mbox\small obs}$ or the Fmax (SI Figure 4); i.e.
the designed riboswitch was constitutively on.

\section{DISCUSSION}
In this paper, we have demonstrated the success of a purely 
computational approach
for the rational design of artificial type III hammerhead ribozymes. 
%Figure~\ref{fig:cleavageHH1-HH10}, 
Figure~4 clearly shows the Mg$^{++}$-dependent cleavage of each designed
sequence HH1-HH10, as well as the non-cleavage of the 8G mutant of each 
sequence, strongly suggesting that cleavage is due to
the usual hammerhead mechanism. 
Cleavage time series data for three to five technical replicates for 
each of the
ten computationally designed hammerheads, displayed in 
%Figure~\ref{fig:kineticsHH1_HH7}
Figure~5 and SI Figure 2,
lead to observed cleavage rates varying 100-fold from
0.0027 min$^{-1}$ for HH3, to 0.25 min$^{-1}$ for HH7.
The relatively fast cleavage rate of HH7, selected from over one million
sequences returned by {\tt RNAiFold} solely on the criteria of
minimizing ensemble defect, {\em with} the additional requirement of
having GUC at the cleavage site,
is slower only by a factor of 10 from wild type hammerhead cleavage
rates (recall that wild type cleavage rates vary between
0.5 and 2 per minute \cite{ClouetdOrval.b97}).
In contrast, HH8 had an observed cleavage rate of 0.02 min$^{-1}$,
although it was selected  solely on the criteria of
minimizing ensemble defect -- {\em without} the additional requirement of
having GUC at the cleavage site. This experimental result suggests that
cleavage kinetics may be the underlying reason that cytidine is present
at cleavage position 8 in 95\% of the 84 sequences in the Rfam seed 
alignment of family RF00008. 
%Why is the presence of cytidine, rather
%than adenine, at the cleavage position 8 presumably so important? Why
%is cytidine found at the cleavage site of 95\% of the 84 sequences in
%the Rfam seed alignment of family RF00008? Unfortunately, the results of
%the current paper do not shed any light on these provocative questions.

Among more than 20 computational features, the features found to be
most highly correlated with cleavage rate
$K_{\mbox\small obs}$ for HH1-HH10 were 
(full) average structural positional entropy,
ensemble defect, and expected base pair distance
discrepancy for `conserved site' with values respectively of
-0.461, -0.370, -0.438. However, this result is based on
a tiny set of data and can only be taken as a suggestive
first step towards a more systematic determination of which
measures of structural diversity/flexibility/rigidity might
best predict ribozyme activity. 

In the design phase, we selected HH1-HH5 to have a positional
entropy profile similar to that of wild type PLMVd, i.e. to have
small average (structural positional) entropy of conserved site, 
based on the intuition that certain positions in the wild type hammerhead
may have high entropy to support cleavage. However, it is presently unclear
whether discrepancy measures (absolute difference between wild type
and synthetic) restricted to the conserved site are useful at all.
Indeed, among all sequences returned by {\tt RNAiFold}, HH6 had an observed
cleavage rate of 0.102/min, a bit less than half that of HH7, yet
HH6 was selected to have the
{\em largest} entropy discrepancy from the conserved site among all
sequences, such that the probability of the MFE structure exceeded
40\%. Without additional experiments on a large collection of 
computationally designed hammerheads, and perhaps without extensive
molecular dynamics modeling, it remains unclear
to what extent hammerhead efficiency, as assayed by cleavage kinetics,
is dependent on matching the positional stability and flexibility of
the wild type PLMVd hammerhead.  
%Using the new RNA folding kinetics algorithm {\tt FFTmfpt}, which is
%part of the software suite {\tt Hermes} under development in our lab,
%the Pearson correlation is 0.696 -- i.e. if predicted folding time
%from the empty structure to the target structure is small then cleavage
%rate is small. This result might seem to be counter-intuitive --
%clearly, more work is required with a substantially larger set of
%computationally designed hammerheads.

%Removed by Peter, following Referee 3
%This is actually good news, since it suggests that the molecular
%structure alone may suffice for synthetic design -- i.e. that 
%positional flexibility/stability may not be of great importance for
%hammerhead function.

It is interesting to note that HH1-HH6 are not recognized as
hammerheads by the Rfam web server \cite{Gardner.nar11}, which
relies on the program {\tt Infernal} \cite{Nawrocki.b09}, a
sophisticated machine learning algorithm (stochastic context free grammar)
that depends on recurring sequence and structural motifs.
Rfam predicts only HH7-HH10 to be type III hammerheads,
with the following confidence scores:
HH7 41.3 bits   (E-value 5.9e-09), HH8 38.1 bits (E-value 4.6e-08),
HH9 37.5 bits (E-value 6.8e-08), HH10 38.9 bits  (E-value 2.9e-08).

Currently, {\tt NUPACK-DESIGN} \cite{Zadeh.jcc11}
appears to be one of the most efficient
tools to design RNAs by employing a heuristic computational
search to minimize ensemble defect.
Given the constraints for synthetic hammerhead design described in this
paper, the {\tt NUPACK} server returned 10 sequences, nine of whose
MFE structures were identical to that of PLMVd AJ005312.1/282-335.
(The {\tt NUPACK} philosophy is that minimizing ensemble defect is more
important than guaranteeing that sequences be an exact solution of the
inverse folding problem. The {\tt NUPACK} 
web server has an upper limit of 10 sequences
that can be returned. In contrast, after downloading and compiling
the NUPACK source code, each run of NUPACK {\tt design} returns a 
single sequence; since the procedure is stochastic, repeated runs will
usually return different sequences.)
The first sequence returned by the {\tt NUPACK} web server
was CGCCGGUAGC
CUGACCCAGG
CCUGAAGAGC
UCUACCCCCC
GAGCGAAACC
GGCU, which has normalized ensemble defect of 2.5\%, the same value as
that of HH8 (1.45179/54 = 0.025030862). The cleavage rate of HH8, whose
cleavage site is GUA (as in the {\tt NUPACK} sequence) is 0.02/min, 
with five faster cleaving synthetic hammerheads. 
Despite the speed of {\tt NUPACK} in designing RNAs with
low ensemble defect, one advantage of 
{\tt RNAiFold} is that prioritization of candidate sequences 
is performed in a postprocessing phase,
thus allowing one to select solutions of inverse folding that are
optimal with respect to various measures (not only ensemble defect), 
as we have done in this paper. 

We have additionally tested the programs {\tt RNAdesign} \cite{RNAdesign}
and {\tt IncaRNAtion} \cite{Reinharz.b13}, with the Rfam consensus structure
of PLMVd hammerhead as target structure. Only 5.84\%	[resp. 2.57\%] of
the sequences returned by {\tt RNAdesign} [resp. {\tt IncaRNAtion}]
actually folded into the target structure, thus requiring substantial
additional computation time to select those sequences that fold into the target
(in constrast, {\tt RNAiFold} returns only sequences that correctly fold
into the target structure). See Supplementary Information and
\url{http://bioinformatics.bc.edu/clotelab/SyntheticHammerheads/}
for comparative results concerning entropy, ensemble defect, etc.

In addition to computationally designing the functional hammerheads
HH1-HH10, we have designed the  166 nt sequence $\seq$, in which
a synthetic hammerhead is embedded within the terminal stem-loop of
the structure depicted in 
Figure~4. 
The sequence $\seq$ is self-cleaving at the expected GUC cleavage site
114-116.
Moreover, as shown in
%Figure~\ref{fig:ribozymeCleavageAssay}
Figure 7D,
cleavage kinetics for this 166 nt artificial ribozyme 
($K_{\mbox\small obs}$=1.3/min) are as fast
as those of wild type hammerheads, although the cleavage amount (Fmax=0.47)
is quite poor compared with our other designed ribozymes HH1-HH10.
By utilizing two mutants, 
one at the cleavage site position 116, and one further downstream at 
position 142 in the CUGUAGA segment necessary for catalysis of
cleavage, we show effectively that cleavage in the
synthetic wild type, designed construct is due to
the usual hammerhead mechanism. Additionally, we have 
demonstrated Mg$^{++}$-dependence, necessary for the cleavage mechanism, 
through the complete absence of 5$'$- and 3$'$-cleavage products when 
incubated for an extended period of time of 
24 hours in buffer lacking Mg$^{++}$.

The software {\tt RNAiFold} solves the inverse folding problem,
not only for a target secondary structure, but as well when the target $S$
is the hybridization of two
secondary structures; i.e. when $S$ contains both intra- and inter-molecular 
base pairs. Since  {\tt RNAiFold} uses constraint programming,
it can perform a complete search of the space of compatible sequences,
and thus return {\em all} sequences, whose MFE structure (resp.
MFE hybridization) is a given target structure (resp. hybridization), or
{\em can certify that no such solution exists}.  
The fact that {\tt RNAiFold} determined that no solution of inverse folding exists
for the GUH to GUG (resp. NUH to GUG) mutant of the target structure depicted in
%Figure~\ref{fig:riboswitchRibozyme}
Figure~3
(resp. the right panel of
%Figure~\ref{fig:nonCleavingStructures}
Figure~6) provides very compelling computational evidence
that there are structural reasons for the reason that GUG is not a hammerhead
cleavage site.

\section{CONCLUSION}

In this paper, by employing our
constraint programming solution {\tt RNAiFold} 
\cite{GarciaMartin13,GarciaMartin.nar13} to generate more than one
million sequences, that agree with 
Peach Latent Mosaic Viroid (PLMVd) AJ005312.1/282-335  at the 15 nucleotides
having greater than 96\% conservation in Rfam RF00008 seed alignment,
and have MFE structure identical to that of the Rfam consensus secondary
structure of PLMVd. Ten candidate hammerheads, which were selected using
criteria that measure either {\em structural diversity} or regional 
{\em structural flexibility/rigidity}, were shown to be functional,
with varying kinetics, by an {\em in vitro} cleavage assay.
This appears to be the first purely computational design 
and experimental validation of novel functional ribozymes. Moreover,
by computationally designing a 166 nt synthetic RNA, whose terminal
stem-loop harbors a functional computationally designed hammerhead,
we show that {\em in silico} design and placement of artificial
hammerheads is possible.

Since {\tt RNAiFold} supports user-defined sequence constraints,
as well as structural compatibility and incompatibility constraints,
our method should be able to rationally design hammerheads that
reside within larger RNAs, which meet user-defined sequence and structure
constraints.

\section{ACKNOWLEDGEMENTS}

We would like to thank Niles Pierce and Brian Wolfe for discussions
concerning {\tt NUPACK}, and to the anonymous referees for their comments.  
Funding for research was provided by the
National Science Foundation with grants DMS-1016618, DBI-1262439.
Any opinions, findings,
and conclusions or recommendations expressed in this material are
those of the authors and do not necessarily reflect the views of the
National Science Foundation.

\subsubsection{Conflict of interest statement.} None declared.

%\begin{footnotesize}
\bibliographystyle{nar}
%\bibliographystyle{plain}
%\bibliography{/Users/clote/text/BIBdir/clote}
%\bibliography{biblio}

%\end{footnotesize}

\input{tables}

\hfill\break\newpage
\input{figures}

\end{document}

%% file: tables.tex
\begin{table*}[b]
\begin{tabular}{|lll|}
\hline
ID	&	Sequence	 & Selection criteria\\
\hline
{\tiny HH1}	& {\tiny UUAAUGUAGAGCGAUUCGUUCCUGAAGAGCUAUAAUUUCUUAGCGAAACAUUAU} & {\tiny GC-content $30-39\%$, $P(S_0,\seq) \geq 40\%$, smallest (binary) entropy distance for conserved site} \\
{\tiny HH2}	& {\tiny UUAUUGUAGCGCGAUUCGCGCCUGAAGAGAUGCGUUUUAACAUCGAAACAGUAU} & {\tiny GC-content $40-49\%$, $P(S_0,\seq) \geq 40\%$, smallest (binary) entropy distance for conserved site} \\
{\tiny HH3}	& {\tiny CUAUUGUAGCGCGAUUCGCGCCUGAAGAGAUCUGUUUUAUGAUCGAAACAGUAU} & {\tiny GC-content $40-49\%$, $P(S_0,\seq) \geq 40\%$, second smallest (binary) entropy distance for conserved site} \\
{\tiny HH4}	& {\tiny UGGAUGUAGCGCGAUUCGCGCCUGAAGAGCGGUCAUCCAUCCGCGAAACAUUCU} & {\tiny GC-content $50-59\%$, $P(S_0,\seq) \geq 40\%$, smallest (binary) entropy distance for conserved site} \\
{\tiny HH5}	& {\tiny CUCAGGUAGCGCGAUUCGCGCCUGAGGAGGGGUCUGGUAUCCCCGAAACCUGAU} & {\tiny GC-content $60-69\%$, $P(S_0,\seq) \geq 40\%$, smallest (binary) entropy distance for conserved site} \\
{\tiny HH6}	& {\tiny UGGCGGUAGCGCGAUUCGCGCCUGAAGAGGGGUAACGCGUCCCCGAAACCGUCU} & {\tiny GC-content $30-39\%$, $P(S_0,\seq) \geq 40\%$, {\em largest} (binary) entropy distance for conserved site} \\
{\tiny HH7}	& {\tiny UCAAUGUCGCGCGAUUCGCGCCUGAAGAGAUGGAAUUUAACAUCGAAACAUUGU} & {\tiny GUC in positions 6-8, smallest ensemble defect}\\
{\tiny HH8}	& {\tiny UCAAUGUAGCGCGAUUCGCGCCUGAAGAGAUGGAAUUUAACAUCGAAACAUUGU} & {\tiny smallest ensemble defect}\\
{\tiny HH9}	& {\tiny UUAAUGUCGCGCGAUUCGCGCCUGAAGAGAUCUGACUUCUGAUCGAAACAUUAU} & {\tiny $P(S_0,\seq) \leq 20\%$, smallest (binary) entropy distance for conserved site}\\
{\tiny HH10}	&	{\tiny UUAAGGUCGCGCGAUUCGCGCCUGACGAGCUAUAUUUUAUUAGCGAAACCUUAU} &	{\tiny smallest (binary) entropy distance for conserved site}\\
\hline
\end{tabular}
\caption{Hammerhead candidates selected and selection criteria used.
Note that, subject to presence or absence of additional constraint C8,  HH7 and HH8 had also the largest probability of structure, the smallest full structural positional entropy, the smallest (Morgan-Higgs and Vienna) structural diversity and smallest expected base pair distance.}
\label{table:tableSequencesChosen}%
\end{table*}

\begin{table*}[b]
\begin{tabular*}{\columnwidth}{@{}lllllll@{}}
\hline
ID	&	K$_{\mbox \small obs}$	&	F$_{\mbox\small \max}$ &	MSE	&	Pos Ent	&	Ens Def	&	EBPD Dis Act\\
\hline
HH1	&	0.037	&	0.79	&	0.0029	&	0.270882	&	4.167687	&	0.0501207	\\
HH2	&	0.0057	&	0.74	&	0.003	&	0.287235	&	4.552053	&	0.0386253	\\
HH3	&	0.0027	&	0.65	&	0.0039	&	0.259577	&	4.121914	&	0.0410984	\\
HH4	&	0.0127	&	0.55	&	0.0048	&	0.403846	&	6.755976	&	0.0354213	\\
HH5	&	0.0085	&	0.52	&	0.0066	&	0.382235	&	6.240083	&	0.033132	\\
HH6	&	0.102	&	0.73	&	0.0047	&	0.414872	&	8.138131	&	0.059864	\\
HH7	&	0.25	&	0.74	&	0.0107	&	0.119159	&	2.383671	&	0.0406728	\\
HH8	&	0.02	&	0.68	&	0.0124	&	0.078518	&	1.45179	&	0.0662421	\\
HH9	&	0.025	&	0.76	&	0.0015	&	0.247886	&	4.525597	&	0.0328018	\\
HH10	&	0.14	&	0.77	&	0.01	&	0.286425	&	4.975979	&	0.0269354	\\
\hline
\end{tabular*}
\caption{Kinetics of cleavage for 10 computationally designed hammerheads, and
correlation with several measures.
Cleavage rate K$_{\mbox \small obs}$ (min$^{-1}$), maximum percent
cleavage F$_{\mbox\small \max}$, mean squared error MSE, 
(full) structural positional entropy
Pos Ent, ensemble defect Ens Def, and expected base pair distance
discrepancy for the `conserved (or active) site' EBPD Dis Act. 
The Pearson correlation
between cleavage rate and Pos Ent, Ens Def, EBPD Disc Active is respectively
-0.461, -0.370, -0.438; i.e. cleavage rate is faster when these secondary
structure deviation values are smaller. 
Other measures, such as structural diversity, had
smaller correlation, while measures such as GC-content and MFE had almost
no correlation with cleavage rate. See Supplementary Information for
full table of correlation for all measures.}
\label{table:kinetics}%
\end{table*}

%% file: figures.tex
%NOTE: Peter replaced \linewidth by \textwidth and Table 3 appeared!

\begin{figure*}
\begin{center}
\includegraphics[width=0.7\textwidth]{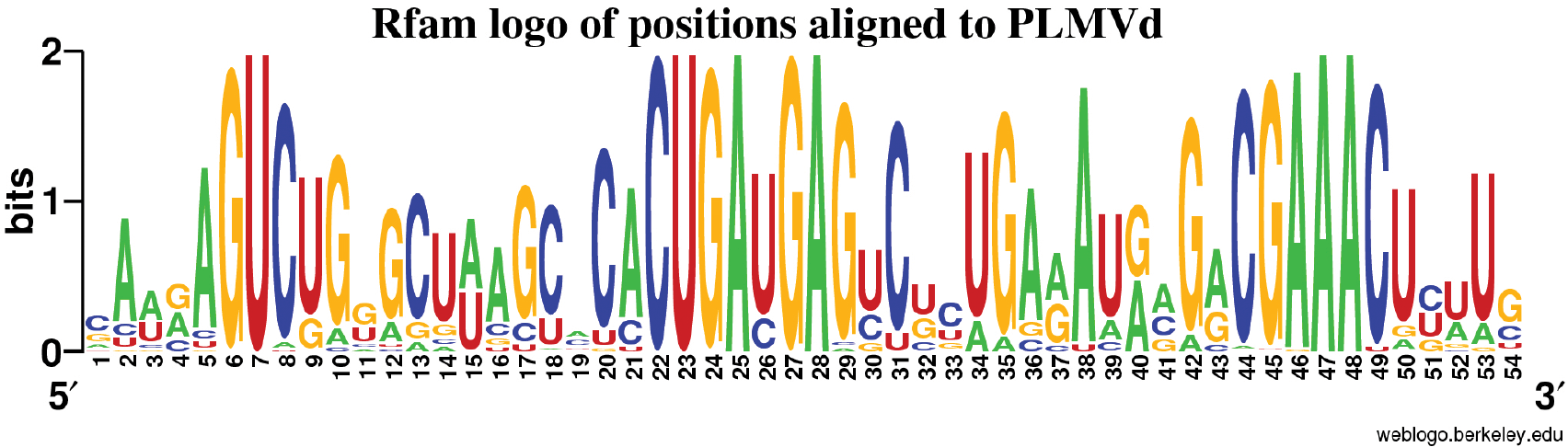}
\includegraphics[width=0.25\textwidth]{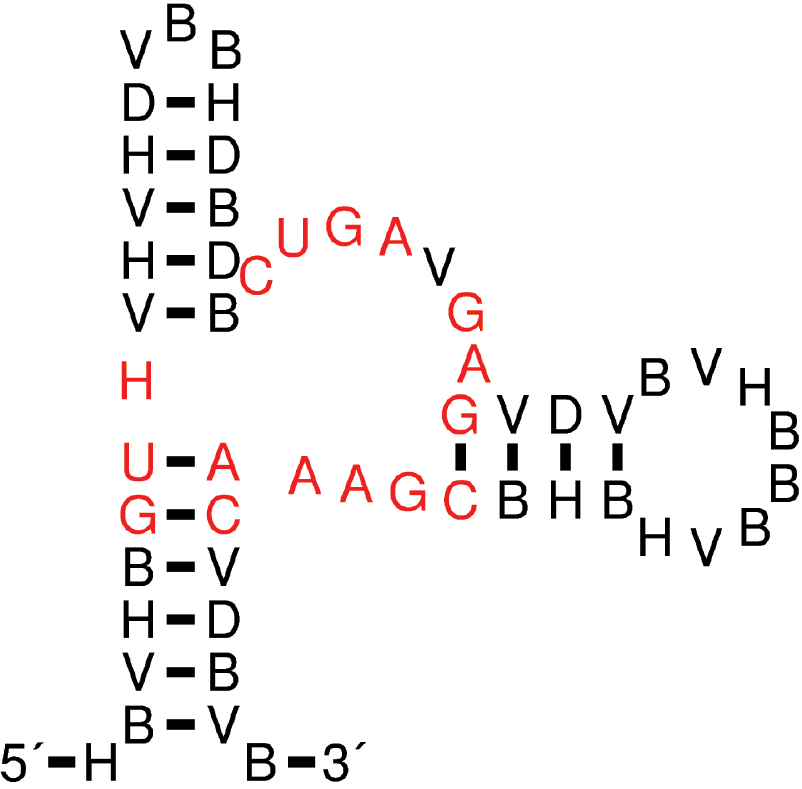}
\end{center}
\caption{ ({\em Left}) Sequence conservation for the 56 nt consensus sequence
for type III hammerhead ribozymes from
version 11.0 of the Rfam database \cite{Gardner.nar11};
image from {\tt http://rfam.sanger.ac.uk/family/RF00008\#tabview=tab3}.
({\em Left}) Sequence logo of conservation at positions aligned with
the 54 nt Peach Latent Mosaic Viroid (PLMVd) AJ005312.1/282-335 from the
hammerhead ribozyme type III seed alignment sequences from Rfam family
RF00008.  In-house program used to determine frequencies of positions 
aligned to those of PLMVd; sequence logo generated with WebLogo 
\cite{Crooks.gr04} (web server at {\tt http://weblogo.berkeley.edu/}).
The 15 positions 6-7, 22-25,27-29, 44-49 of PLMVd had sequence conservation
in excess of 96\%, while cleavage site C at position 8, adjacent to region 6-8, 
was conserved in 94.9367\% of RF00008 seed alignment sequences. {\tt RNAiFold}
was subsequently used to solve the inverse folding problem with consensus
structure of PLMVd used as target, with sequence constraints at positions
6-8, 22-25,27-29, 44-49, as explained in text. Resulting from this analysis,
the sequence constraints for {\tt RNAiFold} were defined to be
{\tt HBVHBGUHVH VHDVBBHDBD BCUGAVGAGV DVBVHBBBVH BHBCGAAACV DBVB}.
({\em Right}) Sequence constraints for {\tt RNAiFold} with indicated
target secondary structure. The 15 positions 6-7, 22-25,27-29, 44-49 having
over 96\% sequence conservation in the seed alignment of RF00008
were constrained to be those in 
Peach Latent Mosaic Viroid (PLMVd) AJ005312.1/282-335, and the cleavage
site 8 was constrained to be H (not G). All 38 remaining
positions were constrained to be distinct from
the corresponding nucleotides in PLMVd.}
\label{fig:PLMVdSequenceLogo}
%Figure1
\end{figure*}

\begin{figure*}
\includegraphics[width=0.9\textwidth]{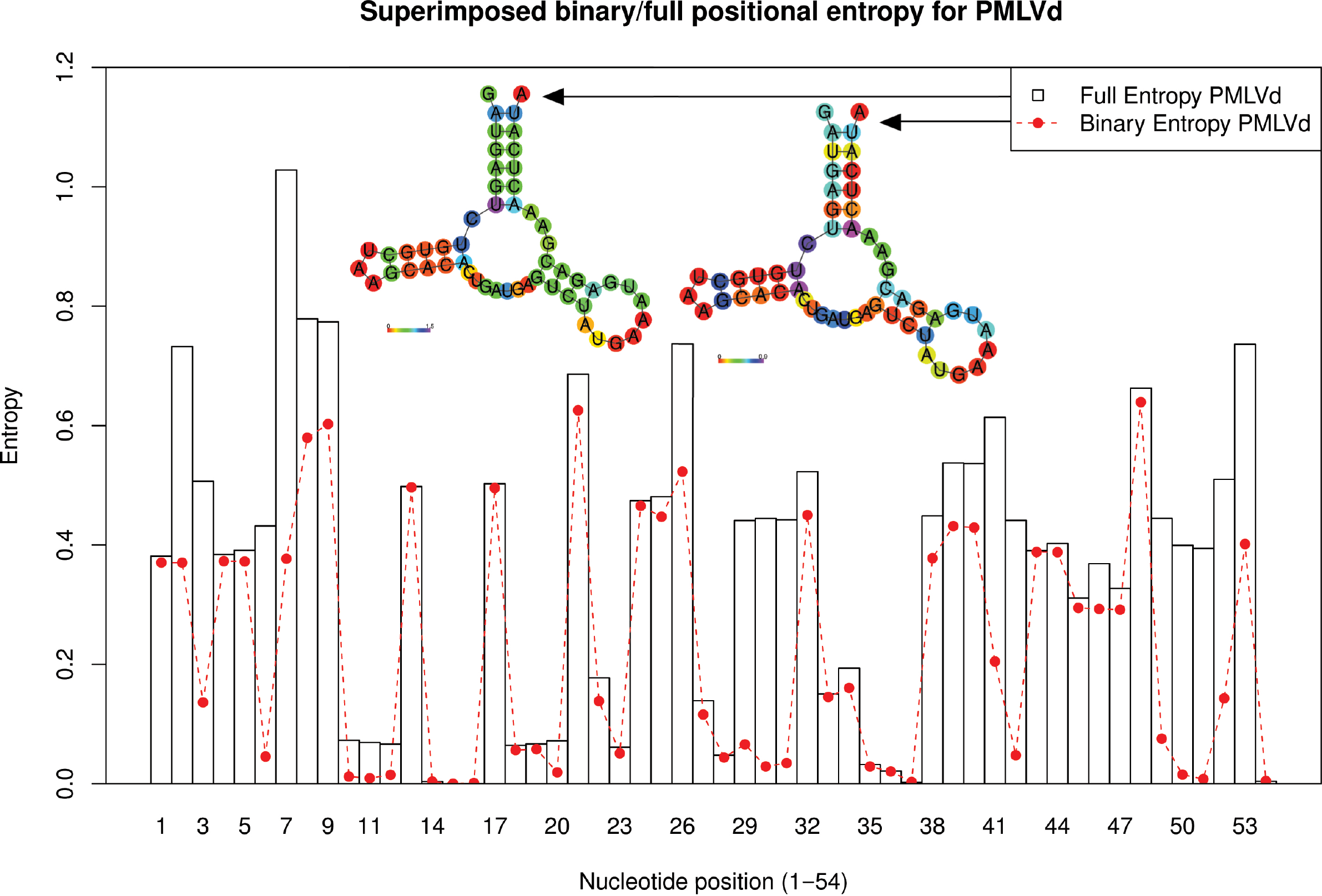}
\caption{Binary and full structural positional entropy 
of hammerhead 
Peach Latent Mosaic Viroid (PLMVd) AJ005312.1/282-335.
({\em Left}) Full structural positional entropy $H$.
({\em Right}) Binary structural positional entropy $H_b$. Note that
positions 50, 51 of have medium (full) entropy and high binary entropy,
which indicates that these positions tend always to be base-paired in
the low energy ensemble of structures, though with different base pairing
partners.  Note that the conserved region
GUH in 6-8 has moderate to high entropy (G6: 0.62, U7: 1.48, H8: 1.12),
GUC in 22-24 has low entropy (G22: 0.26, U23: 0.09, C24: 0.68), 
GAG in 27-29 has low entropy (binary entropy is very low)
(G27: 0.12, A28: 0.04, G29: 0.07), 
while 44-49 has medium entropy.
Left colored secondary structure figure created 
using {\tt relplot.pl} from Vienna RNA Package
\cite{Gruber08}; right upper colored secondary structure
figure created by modifying code {\tt relplot.pl}.
}
\label{fig:superimposedBinaryFullEntropyPlotHHwildtype}
%Figure2
\end{figure*}

\begin{figure*}[!ht]
\begin{center}
\includegraphics[width=0.7\textwidth]{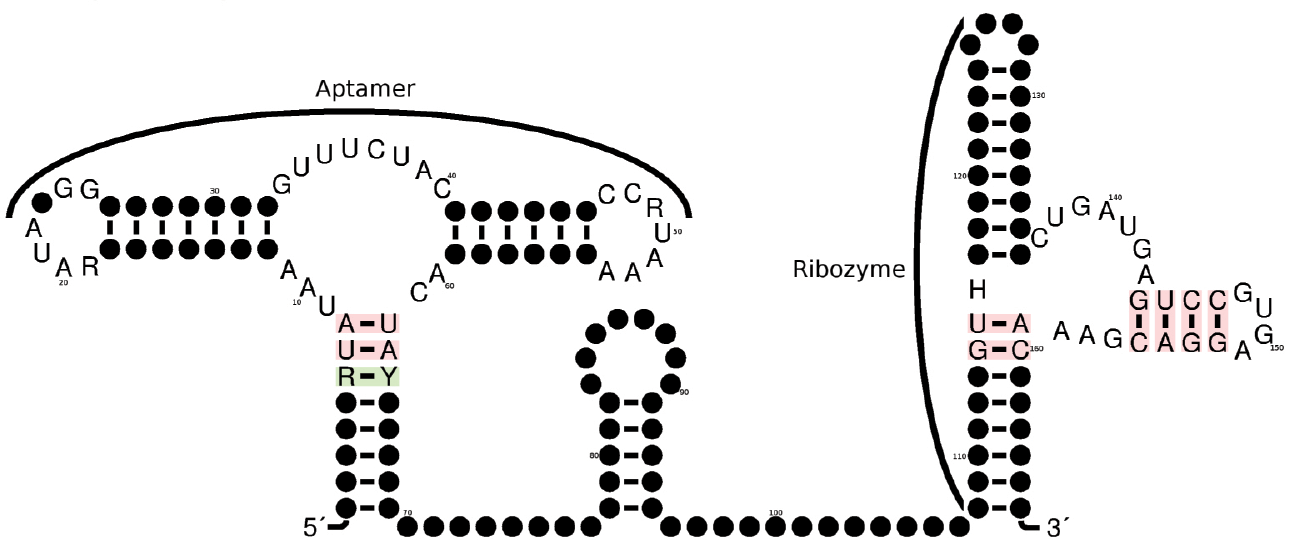}
\caption{Target secondary structure $S$ for modular placement of artificial
hammerhead within larger RNA molecule. The structure and highly conserved
nucleotides (sequence constraints) of the XPT-riboswitch appear on the 
left, while the structure and highly conserved nucleotides of the type 
III hammerhead ribozyme appear on the right. 
}
\label{fig:riboswitchRibozyme}
%Figure3
\end{center}
\end{figure*}

\begin{figure*}
\begin{center}
\includegraphics[width=0.8\textwidth]{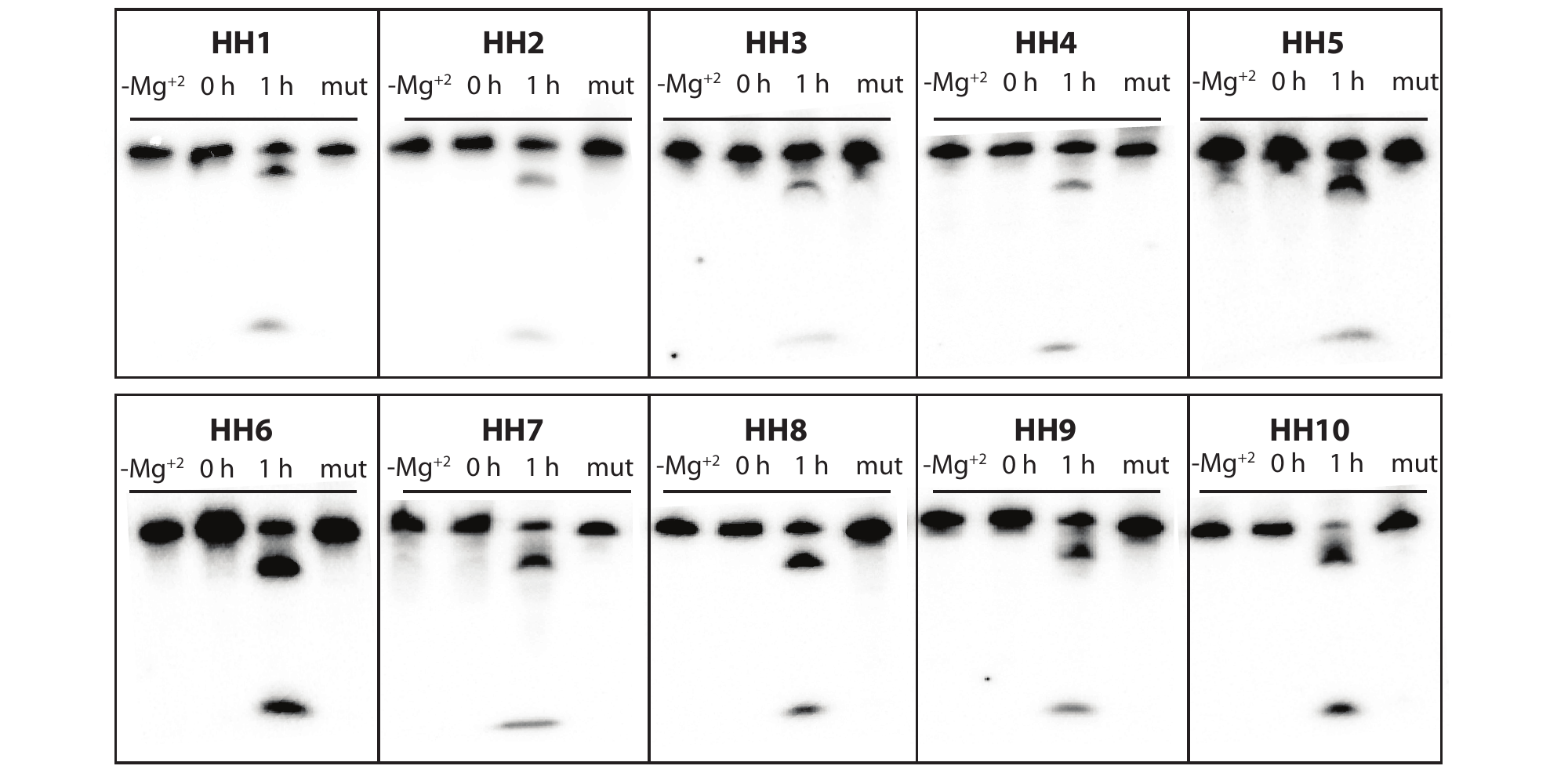}
\end{center}
\caption{Summary of designed hammerhead cleavage. Each designed 
hammerhead RNA was incubated under mild conditions for 1 hour as 
described in the Methods to assess cleavage. As negative controls,
a no magnesium, and a 0 hour reaction were also conducted for each RNA. 
Additionally, the 8G mutation, predicted to be incompatible with the 
hammerhead structure (see Methods), was constructed for each 
designed sequence and examined under equivalent conditions to 
confirm that self-cleavage occurs using the expected hammerhead mechanism.
}
\label{fig:cleavageHH1-HH10}
%Figure4
\end{figure*}

\begin{figure*}[!ht]
\begin{center}
\includegraphics[width=0.9\textwidth]{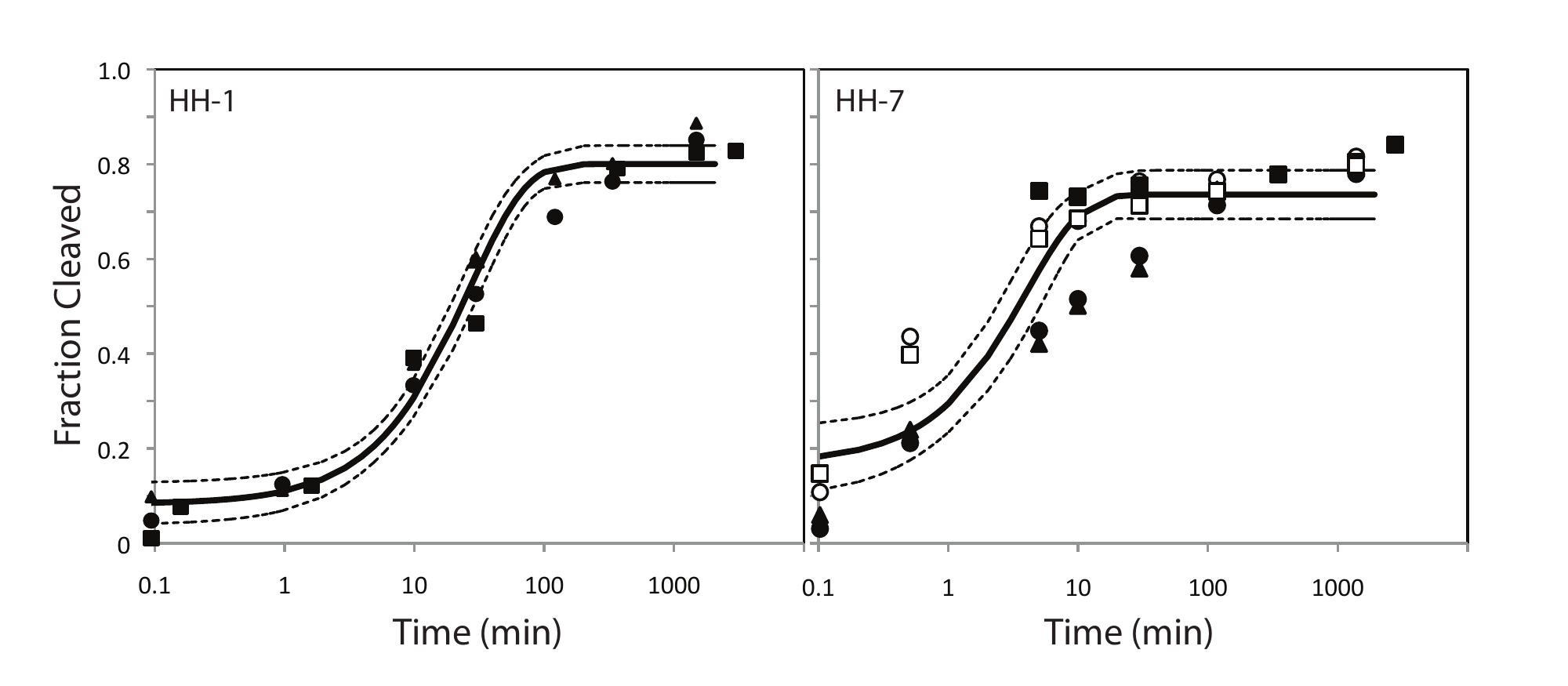}
\caption{{\em (Left)} HH1: typical cleavage time series curve with good
error parameters (standard deviation <10\% of mean, with
mean squared error (MSE) = 0.0029).  Solid line represents fitted line,
and dotted lines indicate 95\% confidence interval. Different
data sets represented by filled and unfilled squares, triangles, etc.
{\em (Right)} HH7: fastest hammerhead cleavage rate, though
determined with considerable error (MSE=0.01).
In data from the first experiments for HH7, indicated by filled squares,
cleavage had been measured at times when maximum cleavage had nearly
occurred (these points appear in the flat part of the fitted curve).
Subsequent datasets have focused on shorter time periods. 
This curve was fitted using five data sets.
Time series curves for cleavage data for the remaining 8 designed
hammerheads HH2-HH6 and HH8-HH10 are shown in Supplementary Information.
}
\label{fig:kineticsHH1_HH7}
%Figure5
\end{center}
\end{figure*}

\begin{figure*}
\begin{center}
\includegraphics[scale=0.8]{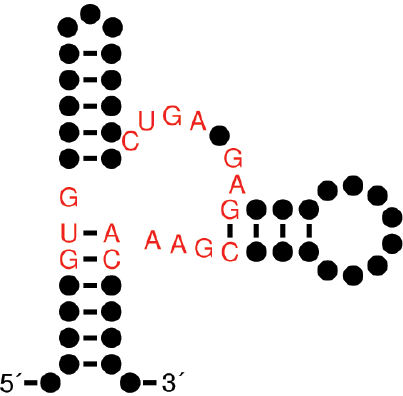}
\hspace{1cm}
\includegraphics[scale=0.8]{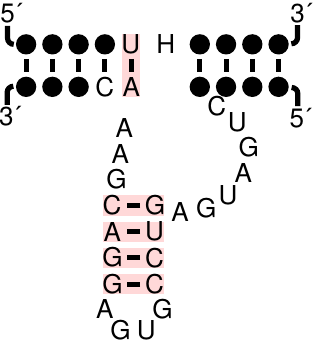}
\end{center}
\caption{{\em (Left)} Target structure $S$ used in computational
experiment with {\tt RNAiFold}, which determined that
{\em no sequence exists}, having guanine at the cleavage site 8
along with those 15 nucleotides of Peach Latent Mosaic Viroid  (PLMVd)
AJ005312.1/282-335 having sequence conservation exceeding 96\%, and which
the Rfam consensus structure of PLMVd (i.e. whose {\tt RNAfold} 1.8.5
MFE structure is the consensus secondary structure of of PLMVd).
{\em (Right)} 
Hammerhead ribozyme (lower molecule)
hybridized with {\em trans}-cleavage target RNA (upper molecule). 
Cleavage site NUH occurs at position 4-6 of
the upper molecule, where `H' denotes `not G'. 
{\tt RNAiFold} shows that no two
sequences $\seq_1,\seq_2$ exist, where $\seq_1$ contains `GUG' at positions 
4-6, both $\seq_1,\seq_2$ contain the other indicated nucleotides, for which
the indicated structure is the MFE hybridization of $\seq_1,\seq_2$.
The nonexistence, as determined by {\tt RNAiFold},
of any sequence folding into target structure $S$, which
has GUG at the cleavage site and satisfies certain additional minimal
constraints, strongly suggests that GUG is not a hammerhead cleavage
site is due to the inability of the molecule to fold into a structure
necessary for nucleophilic attack.
Image of right panel adapted from Figure 3A from \cite{James11}, and
both images produced by R2R \cite{Weinberg.bb11}.
}
\label{fig:nonCleavingStructures}
%Figure6
\end{figure*}

\begin{figure*}[!ht]
\begin{center}
\includegraphics[width=0.9\textwidth]{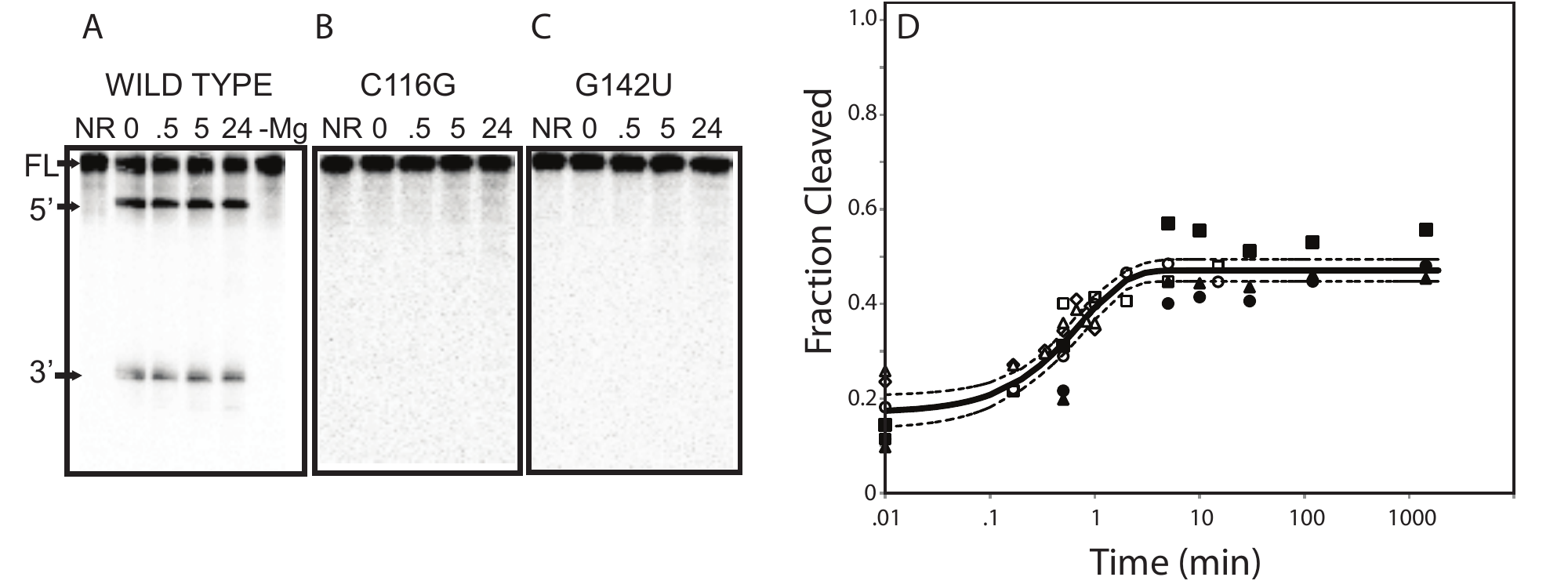}
\caption{ {\em (Left)}
%Peter hours
%Cleavage assay reactions (A,B,C) of designed hammerhead (wild type), mutant C116G, and mutant G142U.  For the wild type (A), mutant C116G (B), and mutant G142U (C) gel images, lane 1 is the undigested RNA (full-length, FT), lanes 2-5 are reactions in cleavage buffer (50 mM Tris pH 7.5, 5 mM MgCl$_2$) at the 0, 30 minutes, 5 hours, and 24 hour time points respectively (5$'$ and 3$'$ cleavage products indicated).  For the wild type (A), lane 6 is a reaction lacking Mg (50 mM tris pH 7.5) incubated for 24 hours. It is evident that cleavage only occurs for the wild type sequence, and when Mg is present.
Cleavage assay reactions (A,B,C) of designed hammerhead (wild type), mutant C116G, and mutant G142U.  For the wild type (A), mutant C116G (B), and mutant G142U (C) gel images, lane 1 is the undigested RNA (full-length, FT), lanes 2-5 are reactions in cleavage buffer (50 mM Tris pH 7.5, 5 mM MgCl$_2$) at the 0 s, 30 min, 5 h, and 24 h time points respectively (5$'$ and 3$'$ cleavage products indicated).  For the wild type (A), lane 6 is a reaction lacking Mg (50 mM tris pH 7.5) incubated for 24 h. It is evident that cleavage only occurs for the wild type sequence, and when Mg is present.
{\em (Right)}
%Peter hour
%Cleavage time series curve (D) for the 166 nt designed hammerhead, with observed cleavage rate of 1.3/min with an Fmax of 0.47 and MSE of 0.0026. This construct displays kinetics comparable with that of wild type hammerheads, although the cleavage amount Fmax is much lower than that of wild type hammerheads.  }
Cleavage time series curve (D) for the 166 nt designed hammerhead, with observed cleavage rate of 1.3/min with an Fmax of 0.47 and MSE of 0.0026. This construct displays kinetics comparable with that of wild type hammerheads, although the cleavage amount Fmax is much lower than that of wild type hammerheads.  }
\label{fig:ribozymeCleavageAssay}
%Figure7
\end{center}
\end{figure*}